# High-Speed Area-Efficient Hardware Architecture for the Efficient Detection of Faults in a Bit-Parallel Multiplier Utilizing the Polynomial Basis of GF($2^m$)

Saeideh Nabipour
Department Computer and Electrical Engineering
University of Mohaghegh Ardabili
Ardabil, IRAN
Saeideh.nabipour@gmail.com

Javad Javidan
Department Computer and Electrical Engineering
University of Mohaghegh Ardabili
Ardabil, IRAN
javidan@uma.ac.ir

*Abstract*—The utilization of finite field multipliers is pervasive in contemporary digital systems, with hardware implementation for bit parallel operation often necessitating millions of logic gates. However, various digital design issues, whether inherent or stemming from soft errors, can result in gate malfunction, ultimately can cause gates to malfunction, which in turn results in incorrect multiplier outputs. Thus, to prevent susceptibility to error, it is imperative to employ a reliable finite field multiplier implementation that boasts a robust fault detection capability. In order to achieve the best fault detection performance for finite field detection performance for finite field multipliers while maintaining a low-complexity implementation, this study proposes a novel fault detection scheme for a recent bit-parallel polynomial basis over GF($2^m$). The primary concept behind the proposed approach is centered on the implementation of an efficient BCH decoder that utilize Berlekamp-Rumsey-Solomon (BRS) algorithm and Chien-search method to effectively locate errors with minimal delay. The results of our synthesis indicate that our proposed error detection and correction architecture for a 45-bit multiplier with 5-bit errors achieves a 37% and 49% reduction in critical path delay compared to existing designs. Furthermore, a 45-bit multiplicand with five errors has hardware complexity that is only 80%, which is significantly less complex than the most advanced BCH-based fault recognition techniques, such as TMR, Hamming's single error correction, and LDPC-based methods for finite field multiplication which is desirable in constrained applications, such as smart cards, IoT devices, and implantable medical devices.

## I. INTRODUCTION

Field-level product dependability, such as that of memory and logic blocks, has become a major concern, especially as technology has advanced. As the density of integrated circuits continues to rise, the susceptibility of high-performance integrated circuits to a variety of malfunctions will increase. These circuits are renowned for their high operating frequencies, low voltage levels, and minimal noise margins, which render them even more vulnerable [17]. important areas including error-detecting codes (EDC), coding theory, computational algebra, quasi-random number generation, and digital signal processing are convergent towards finite field arithmetic. In the literature, multiplication has received the most attention among the fundamental arithmetic operations over finite fields GF($2^m$) [2], [16], [18], [21]. This is mainly because it is more complicated than a finite field adders, and by repeatedly performing a multiplication operation, one can execute more difficult field operations like inversion and exponentiation [1], [2]. Numerous error detection techniques for finite field multipliers have been investigated in the literature. In [14], a parity prediction approach has been proposed to detect errors in bit-serial multipliers in GF($2^m$). Their proposed method defines a specific class of fields for error detection, utilizing irreducible all-one polynomials. In [3], [9], the parity prediction technique has been applied for detecting faults in the architecture of the Advanced Encryption Standard (AES). The Advanced Encryption Standard (AES) facilitates the encryption and decryption of a data block through the utilization of a block key. Triple Modular Redundancy (TMR) serves as an instance of a space redundant scheme [8]. In [9], a methodology predicated upon space redundancy has been advanced, where the functional block is replicated on multiple occasions and the accuracy of the output is assessed via a voter. Error detection and recalculation techniques represent well-established fault tolerant schemes for identifying/correcting individual errors, which were presented in [4] and [10]. To address the issue of double error detection and single error correction, various methods, such as the SEC/DED schemes have been proposed. The Hamming codes and LDPC codes have been suggested as the basis for a single error correction (SEC) scheme in [6] and [7], respectively. In [10] and [11], a novel dynamic error correction architecture was proposed utilizing BCH codes for bit parallel polynomial basis (PB) multiplication.

In this study, the proposed scheme is implemented by the an efficient BCH fault detection method presented in [10] and [11]. Our objective is to examine the fault detection architecture for a novel bit-parallel polynomial basis (PB) multiplier GF($2^m$) [17]. The architecture is based on an area-delay-efficient BCH decoder architecture that can detect and correct multiple errors. Our proposed technique differs fundamentally from the BCH based fault detection technique investigating in [11] and [12] in two critical aspects. First and foremost, our innovative method for a bit parallel polynomial basis (PB) multiplier achieves exceptional reductions in area, rendering it particularly conducive to situations where space complexity and energy preservation take precedence over time complexity. For achieving a more efficient area-delay implementation of our BCH decoder, the use of the re-encoding technique and FIBM algorithm have been incorporated in BCH decoder. Moreover, we have utilized the Berlekamp-Rumsey-Solomon (BRS) algorithm [19], [22] in conjunction with the Chien-search method to effectively identify the position of errors with least delay. we have demonstrated that the proposed algorithms require less time and area overhead when compared to other methods previously proposed in the literature. The structure of this paper is organized in the following manner: In Section II, we present notations regarding Galois field (GF) arithmetic and a novel bit-parallel polynomial basis (PB) multipliers in GF($2^m$). In Section III, we introduce a fault detection architecture for the proposed bit-parallel multipliers based on BCH codes. Section IV outlines the results of our experiments. Finally, in Section V, we provide concluding remarks.

## II. FUNDAMENTALS OF GALOIS FIELD (GF) ARITHMETIC

A Galois field can be entirely constructed via an $m^{th}$ degree monic polynomial of the form over $GF(2^m)$, $f(X) = x^m + \sum_{j=m-1}^{1} f_j x^j + 1$ is assumed to be an irreducible polynomial. The irreducibility of a polynomial over GF($2^m$) of degree $m$ can be established if it is not divisible by any polynomial over GF($2^m$) of degree less than $m$ [15]. In polynomial basis, all elements of the extension field $GF(2^m)$ can be formed as a result of polynomials over GF(2) of degree less than $m$ [15]. Assume $\alpha \in GF(2^m)$ shows a root of the irreducible polynomial over $GF(2^m)$. Then the production of the polynomial basis is facilitated through a set of m elements, $(1, \alpha, \alpha^2, ..., \alpha^{m-1})$, and $f(\alpha) = 0$. In GF($2^m$), The standard basis multiplication in GF($2^m$) typically comprises of two integral steps: the ordinary polynomial multiplication and field polynomial reduction. Assume $A(x) = a_0 + a_1 x + ... + a_{m-1} x^{m-1}$ and $B(x) = b_0 + b_1 x + ... + b_{m-1} x^{m-1}$ be two field elements and $C(x) = c_0 + c_1 x + ... + c_{m-1} x^{m-1}$ their product module where all $a_j, b_j, c_j \in GF(2)$, the finite field multiplication can be executed via the implementation of the ensuing equation [2]:

$$C(x) = A(x) \times B(x) \mod f(x)$$
$$= ((a_0 + a_1 x + ......... + a_m x^{m-1}) \times (b_0 + b_1 x + ....... + b_m x^{m-1})) \mod f(x)$$
$$= \sum_{j=0}^{m-1} b_j x^j A(x) \mod f(x) \quad (1)$$

In this section, a recently proposed class of low complexity multipliers based on the method of serial interleaved multiplication [18] is examined. The proposed algorithm utilizes a logical relation to replace logical XOR (XNOR) gates with NAND gates, resulting in more efficient multiplication implementation in terms of hardware complexity. This is primarily due to NAND gates having lower area and time complexities compared to other gates, such as AND or XOR/XNOR, which can significantly improve hardware complexity. Our work adopts the same multiplier structure as that of [18]. For the completeness of this paper, we provide a brief formulation of the PB bit parallel multiplier as well. Let $A$ and $B$ are the two multiplicands, to obtain the polynomial with a degree of no more than $2m-2$, the product $A(x) \times B(x)$ in equation (1) must be computed initially. The subsequent stage involves modular reduction, which yields the polynomial $C(x)$ with a degree of at most $m-1$. It is possible to perform recursive computation of each $b_j x^i A(x)$ in equation (1) as follow:

$$C(x) = ...((b_{m-1} x A(x) \mod f(x) + b_{m-2} A(x)) x \mod f(x) +$$
$$b_{m-3} A(x)) x \mod f(x) + ... + b_1 x A(x) \mod f(x) + b_0 A(x) \quad (2)$$

The summation is executed over the range of $I = 0, 1,...,m-1$ starting from 0 to $m-1$, in $m$ iterations. It is noteworthy that the modular reduction stage computes $xA(x)$, which is subsequently multiplied by $b_i$. Consequently, by interleaving $f(x)$ across equation (2), the ensuing equation can be expressed as follows:

$$P^{(1)}(x) x \mod f(x)$$
$$= (p_{m-1}^{(1)} x^m + p_{m-2}^{(1)} x^{m-1} + ... + p_1^{(1)} x + p_0^{(1)}) \mod$$
$$(f_{m-1} x^{m-1} + f_{m-2} x^{m-2} + ... + f_1 x^1 + f_0)$$
$$= (p_{m-1}^{(1)} f_{m-1} + p_{m-2}^{(1)}) x^{m-1} + (p_{m-1}^{(1)} f_{m-2} + p_{m-3}^{(1)}) x^{m-2}$$
$$+ ... + (p_{m-1}^{(1)} f_1 + p_0^{(1)}) x + p_{m-1}^{(1)} \quad (3)$$

The term of $b_{m-1} A(x)$ should be iteratively executed for $(m-1)$ times and may be regarded as $P^{k-1}(x)$. Therefore, a modulo reduction is performed on $xp^{k-1}(x)$, where $p^{k-1}(x)$ represents a partial-product polynomial that is formulated during the iteration process of $(k-1)^{th}$. It can be expressed as a polynomial of $(m-1)$ degree format:

$$P^K(x) = P^{(k-1)}(x) + b_{m-k} A(x) \quad (4)$$

For each $k \leq m$, the equation (4) can be evaluated as follows, which $k$ is the number of iterations:

$$for \quad k = 1: \quad p^{(0)}(x) = 0$$
$$for \quad k = 2: \quad P^{(1)}(x) = b_{m-1} A(x) \quad (5)$$

At the first, for evaluating $P^1(x) x \mod f(x)$, $P^1(x)x$ as the product polynomial needs to be performed and then be reduced using $f(x)$ as follow:

$$C(x) = ...((b_{m-1} x A(x) + b_{m-2} A(x)) x + b_{m-3} A(x)) x + ... +$$
$$b_1 x A(x) + b_0 A(x)) \mod f(x) \quad (6)$$

Based on equation (6), it is evident that the process of modular reduction can be simplified to the summation of variables $p_{m-1}^{(k-1)} f$ and $P.x^i$. By performing a left shift on

variable $P$ for $i$ iterations, the value of $P.x^i$ can be accurately computed as follow:

$$P^{(K-1)}(x) = p_{m-1}^{(k-1)} f + P.x^i \quad (7)$$

The inclusion of the product term $b_{m-2}A(x)$ into equation (2) is necessary on account of the left-most term in equation (1), to achieve the ultimate multiplication outcome of equation (1). It is imperative that the equation $P^K(x) = P^{(k-1)}(x) + b_{m-k}A(x)$ is replicated $m$ times to facilitate the ultimate multiplication result of equation (1). It is noteworthy that the XOR operator is employed to carry out the summation. Consequently, equation (7) can be expressed as:

$$P^K(x) = P^{(k-1)}(x) \oplus b_{m-k}A(x) \quad (8)$$

Based on logical equation of (a XOR b) = ((a NAND (a NAND b)) NAND ((a NAND b) NAND b)), XOR gates can be reduced to a circuit that employs only NAND gates. The circuit diagram for this implementation is presented as follow:

$$P^K(x) = P^{(k-1)}(x) \oplus (b_{m-k}.A(x))$$
$$= P^{(k-1)}(x).\overline{b_{m-k}A(x)} + \overline{P^{(k-1)}(x)}.b_{m-k}A(x)$$
$$= P^{(k-1)}(x).\overline{[P^{(k-1)}(x) + \overline{b_{m-k}A(x)}]} +$$
$$\overline{[P^{(k-1)}(x) + \overline{b_{m-k}A(x)}]}.b_{m-k}A(x)] \quad (9)$$
$$= \overline{P^{(k-1)}(x)[\overline{P^{(k-1)}(x).b_{m-k}.A(x)}].[\overline{P^{(k-1)}(x).b_{m-k}.A(x)}.(b_{m-k}A(x))}$$
$$= \overline{P^{(k-1)}(x) \wedge [\overline{P^{(k-1)}(x) \wedge b_{m-k} \wedge A(x)}] \wedge}$$
$$[\overline{P^{(k-1)}(x) \wedge b_{m-k} \wedge A(x)}] \wedge (b_{m-k}A(x))$$

The logical NAND operation ($\wedge$), can be employed instead of the logical XOR operation to achieve the requisite multiplication outcome, as evidenced in equation (9). This approach has the potential to significantly reduce hardware overhead. The suggested design for the sequential PB multiplier's architecture is depicted in Figure 1. Additionally, the hardware details for the proposed design are explicated in Figure 2. Both modules are constructed with three levels of logic gates, each of which encompasses an array of $m$ NAND gates to execute the NAND operations suggested in equation (9). For more details refer to reference [18].

## III. PROPOSED FAULT TOLERANT MULTIPLIER

### 3.1. BCH Codes

In this section, we propose a novel technique for designing fault-tolerant bit-parallel multipliers over GF($2^m$) utilizing BCH error correction codes. Fig. 3 illustrates the architecture of the suggested fault-tolerant multiplier, which includes a single encoder and decoder block in addition to the functional unit of the multiplier.

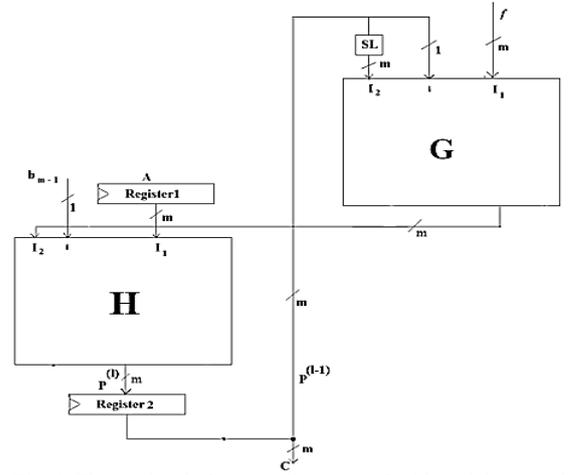

Fig. 1. The top-level of proposed structure for bit-serial sequential multiplier [18]

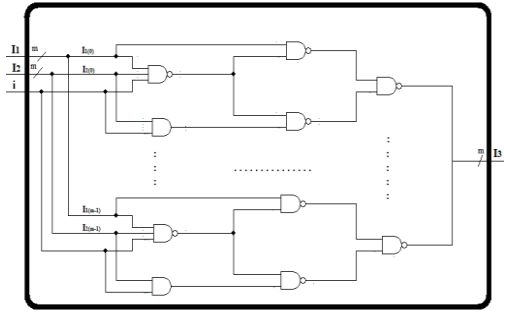

Fig. 2. Logic diagram of module G and H [18]

As depicted in Fig. 3, the multiplier block's output is regarded as a message that is encoded through the BCH encoder prior to being transmitted to the error detection and correction block. Recently, the subject of BCH codes has gained a significant amount of attention due to their remarkable capacity to rectify and identify multiple errors. It is noteworthy that for any pair of positive integers $m(m \geq 3)$ and $(t < 2^{m-1})$, a binary BCH code can be found that is characterized by a block length $n = 2^m - 1$, the number of information bits $k \geq n - mt$, and minimum distance $d_{min} \geq 2t+1$. This code has the capability to correct any combination of $t$ or fewer errors within a block consisting of a certain number of digits [15], [20].

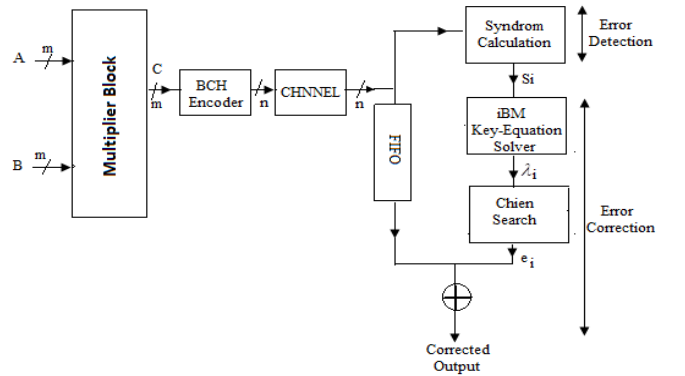

Fig.3. BCH fault-tolerant multiplier architecture

## 3.2. BCH Decoder

Fig. 4 shows a block diagram of the decoding process for a *t*-error correcting BCH code, including three basic steps: syndrome calculation, error locator polynomial calculation, and the Chien search algorithm for locating the errors.

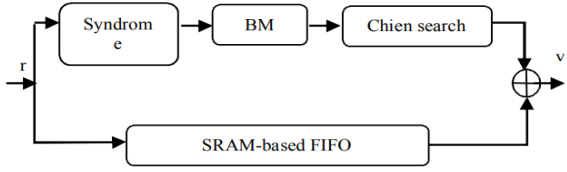

Fig. 4. BCH decoder design [20]

### A. SYNDROME GENERATOR

The initial phase of decoding a BCH code involves determining errors by calculating the syndrome using the received vector *r(x)* as follow:

$$S_i = r(\alpha^i) = c(\alpha^i) + e(\alpha^i) = \sum_{j=0}^{n-1} e_i (\alpha^i)^j \quad (10)$$

The syndromes are not all zero if the received codeword contains errors. Our syndrome block has been implemented in parallel for low-latency implementation. As pointed out in [15] for binary BCH codes, we have $S_{2j} = S_j^2$, so only *t* parallel syndrome generators are needed to calculate the odd-indexed syndromes, followed by considerably simpler square circuits that can process *p* input bits in one cycle, as mentioned in [15]. A codeword, denoted as *r(x)*, that is obtained from a channel is subjected to re-encoding as a preliminary measure for syndrome computation. This re-encoding approach, as stated in reference [5], reduces the decoding latency of BCH. It's an efficient technique that requires no additional circuitry since it employs the BCH encoder that is idle at that moment. Consequently, a re-encoding strategy is applied to our decoder's syndrome generator before actual decoding, with no extra hardware expenses, by repurposing the BCH encoder that remains inactive at that time. Further information can be found in reference [5].

### B. Finding Error-Location Polynomial:

. If any of the syndrome values are non-zero, there has been an error, and this step is used to obtain error location polynomial. The Berlekamp-Massey algorithm (BMA) [15] is a well-known iterative approach for determining error-location polynomials. The error locator polynomial can be calculated using a variety of hard-decision decoding techniques, including the Peterson algorithm, Berlekamp–Massey algorithm, and the Euclidean algorithm among which Berlekamp-Massey (BM) algorithm is the most well-known algoritm in BCH decoding process. Furthermore, many formulation algorithms of the BM method have been proposed in the literature among which FiBM technique, with its minimal hardware complexity, is the most suitable algorithm for binary BCH code. It is ideal for low-power VLSI implementations, such as the one utilized in our BCH decoder. For a more comprehensive understanding, kindly refer to reference [13].

### C. CHIEN SEARCH ALGORITHM

The Chien search algorithm is the most prominent algorithm, that evaluate all possible *2m* Galois field elements in the error polynomial to detect error position. If the output is equivalent to zero, a root would emerge for the polynomial. The Chien search circuit operates by generating an error vector *e*, employing a sophisticated approach. if $\alpha^i$ is a root, then the *(n−i)th* component $e_{n-i} = 1$; otherwise $e_{n-i} = 0$ for all $0 \leq i \leq n-1$. Finally, error will be corrected in the received codeword.

## 3.3. Improved Error Locator Design

Finding of the positions of the errors is one of the most time-consuming stages of BCH decodeing. The time complexity for the Chien exploration algorithm is exceedingly high, particularly when dealing with extended domains and polynomials with large power values. To enhance the efficiency of the root-finding predicament for BCH decoder, a novel approach is developed. The Berlekamp-Rumsey-Solomon (BRS) algorithm [19], [22], in conjunction with the Chien-search method, is devised to locate the roots of the error locator polynomial for BCH decoder. Prior to expounding upon the algorithm, it is imperative to take into account certain definitions and a theorem that are indispensable to formulate this algorithm.

**Definition 1**: The polynomial denoted by *L(y)* in the finite field of *2m* elements is referred to as a *p*-polynomial or linearized polynomial specifically for the case where *p = 2*.

$$L(y) = \sum c_i y^{2i}, where, \ c_i \in GF(2^m) \quad (11)$$

**Definition 2:** An affine polynomial is a polynomial such as *A(y)* over *GF(2$^m$)* if

$$A(y) = L(y) + \beta \quad (12)$$

Where *L(y)* is a *p*-polynomial and $\beta \in GF(2^m)$.

**Theorem 1**: Let $y \in GF(2^m)$ and let us consider the standard basis, denoted by $\alpha^0, \alpha^1, \alpha^2, ..., \alpha^{m-1}$. If *y* is depicted according to the standard basis, i.e., if

$$y = \sum_{k=0}^{m-1} y_k \alpha^k \quad (13)$$

where $y_k \in GF(2), then$:

$$L(y) = \sum_{k=0}^{m-1} y_k L(\alpha^k) \quad (14)$$

Let us consider a simple quadratic *p*-polynomial over GF(2³), such as $A(y) = y^2 + a^3 y + a^4 = 0$ that can be rewritten as:

$$y^2 + a^3 y = a^4 = \alpha^2 + \alpha + 1 \quad (15)$$

Let *α* be the primitive element of GF(2³). Here, $P(x) = x^3 + x^2 + 1$ is the primitive polynomial. If we consider left-hand side of equation (15) as $L(y) = y^2 + a^3 y$,

which is a *p*-polynomial over $GF(2^3)$, (15) can be expressed as:

$$L(y) = \alpha^2 + \alpha + 1 \quad (16)$$

If $y = y_2\alpha^2 + y_1\alpha + y_0 \in GF(2^3)$, with regard to theorem 1, (16) can be written as:

$$y_2 L(\alpha^2) + y_1 L(\alpha) + y_0 L(\alpha^0) = \alpha^2 + \alpha + 1 \quad (17)$$

Where: 
$$L(\alpha^0) = L(1) = 1 + \alpha^3 = \alpha^2,$$
$$L(\alpha) = L(1) = \alpha^2 + \alpha^3 \cdot \alpha = \alpha + 1, \quad (18)$$
$$L(\alpha^2) = \alpha^4 + \alpha^3 \cdot \alpha^2 = \alpha^2$$

By implementing the process of substitution, the equation (18) can be incorporated into the equation (17), thus:

$$(y_2 + y_0)\alpha^2 + y_1\alpha + y_0 = \alpha^2 + \alpha + 1 \quad (19)$$

It can be expressed as a matrix form as follow:

$$\begin{bmatrix} y_2 & y_1 & y_0 \end{bmatrix} \begin{bmatrix} 1 & 0 & 0 \\ 0 & 1 & 0 \\ 1 & 0 & 1 \end{bmatrix} = \begin{bmatrix} 1 & 1 & 1 \end{bmatrix} \quad (20)$$

It is obvious that the roots of (15) can be found by solving the three simultaneous equations given in (20) with three unknown $y_0$, $y_1$ and $y_2$. Evidently solutions of (19) are y = 011 and y = 110, which are two roots of (21). From this example, it is clear that instead of all of the values $L(\alpha^0), ..., L(\alpha^6)$ required in the Chien-search method, computing of the three values $L(\alpha^0)$, $L(\alpha^1)$ and $L(\alpha^2)$ is only needed. The main feature of this algorithm is grouping of some summands of the polynomial of degree not higher than 11 into multiples of affine polynomials which can be evaluated using very small pre-computed tables leading to fast computation of error locations. for further details, refer to reference [19], [22].

## VI. EXPERIMENTAL RESULT

The proposed fault-tolerant based on BCH error correction code for two structures of 16-bit and 45-bit multiplier has been modeled in VHDL, followed by simulation and synthesis using the Synopsys tools in the UMC technology library, utilizing the 0.18 micron CMOS technology. The evaluation of power consumptions is achieved by using Synopsys Power CompilerTM. It should be noted that this technique can be extended to bit parallel multipliers of any size. The comparison of our method versus the proposed multipliers in [11] has been shown in Table 1. The proposed multiplier implementation in 180nm technology depicates a remarkable 69% and 66% reduction in area-delay product (ADP) and 72% and 76% reduction in power-delay product (PDP) for 16-bit and 45-bit multiplier, respectively, as compared to multiplier in [10]. Moreover, the proposed architecture in 90nm technology offers an outstanding 88% and 98% area-delay improvement (ADP) and 85% and 81% power-delay product improvement (PDP) for 16-bit and 45-bit multiplier, respectively, in comparison with the multiplier in [11]. The FPGA implementation results further validate the superiority of against the multiplier in [11].

Table.1. Comparison of 16-bit and 45-bit GF Multiplier factors with [11].

| Parameters | 16-bit Multiplier [11] | proposed 16-bit Multiplir | 45-bit Multiplier [11] | proposed 45-bit Multiplir |
|---|---|---|---|---|
| Area $_{(\mu^2 m)}$-180nm | 10863.2 | 6523.2 | 77514.5 | 56212 |
| TPower ($\mu W$) 180m | 489 | 264 | 3300 | 1650 |
| Delay(ns)-180nm | 3.11 | 1.6 | 6.86 | 3.2 |
| ADP | 33784.55 | 10437.12 | 531749.47 | 179878.4 |
| PDP | 1520.79 | 422.4 | 22638 | 5280 |
| Reduction in ADP | 0.69% | - | 0.66% | - |
| Reduction in PDP | 0.72% | - | 0.76% | |
| Area $_{(\mu^2 m)}$-90nm | 3029.4 | 1065.2 | 19795.6 | 564.2 |
| TPower ($\mu W$) 90nm | 78.5 | 34.2 | 375.46 | 105.45 |
| Delay(ns)-90nm | 0.6 | 0.2 | 1.06 | 0.7 |
| ADP | 1817.64 | 213.04 | 20983.33 | 394.94 |
| PDP | 47.1 | 6.84 | 397.98 | 73.81 |
| Reduction in ADP | 0.88% | - | 0.98% | - |
| Reduction in PDP | 0.85% | - | 0.81% | - |

Table 2 presents a comparison of the delay overhead for proposed method against [11]. The BCH code used for the comparison is 16-bit and 45-bit has the capability of correcting up to 5 errors. In comparison to the BCH decoder design in [11], the proposed method, which incorporates the Berlekamp-Rumsey-Solomon (BRS) algorithm and Chien-search method, demonstrates a superior computation delay. Specifically, the computation delay for error detection in the multiplier output is 0.56% and 0.66% for BCH(31,16) and 0.37% and 0.49% for BCH(63,45) in 180nm and 90nm technology, respectively. These values are lower than the Chien search block in [11]. Therefore, the proposed algorithm offers the lowest delay in identifying the position of errors in the multiplier output.

Table.2. Delay Comparison of Proposed ECC Blocks BCH with [11].

| Scheme | Delay(ns)-180 nm | Delay(ns)-90 nm | Reduction in Delay-180nm | Reduction in Delay-90nm |
|---|---|---|---|---|
| BCH(31, 16) [11] | 7.8 | 1.95 | 0.56% | 0.66% |
| Our BCH(31, 16) | 3.4 | 0.658 | - | - |
| BCH(63, 45) [11] | 11.76 | 2.37 | 0.37% | 0.49% |
| Our BCH(63, 45) | 7.4 | 1.2 | - | - |

Table 3 presents a comparison of the complexity of the hardware in our innovative fault-tolerant approach with [6], [7], [8], and [11]. The area complexity required for correcting 3-bit errors in a 45-bit multiplier is only 65% of that required by the TMR [8] method, which is 200%.

Compared to the prevailing single Hamming [6] and LDPC [7] techniques, which incur 120% and 130% hardware overhead, respectively, the proposed technique is far superior.

Table.3. comparison of the area overhead of our approach with the 45-bit multiplie in [6], [7], [8] and [11].

| Scheme | #Errors Correction | Coding Technique | Hardware Complexity |
|---|---|---|---|
| [6] | Single | Hamming | >120% |
| [7] | Single | LDPC | >130% |
| [8] | Multiple | Voting | >200% |
| [11] | 3 | BCH | 150.4% |
| [11] | 4 | BCH | 164.04% |
| [11] | 5 | BCH | 170.4% |
| Our Method | 3 | BCH | 65% |
| Our Method | 4 | BCH | 72% |
| Our Method | 5 | BCH | 80% |

Furthermore, the proposed method yields hardware improvements of 85.4%, 92.4%, and 90.4% for 3, 4, and 5 errors, respectively, compared to [11]. It is evident that our fault-tolerant algorithm is more area-efficient. This is mainly because the utilization of the re-encoding method, which is executed prior to the actual decoding and can be performed for the syndrome generator without incurring any additional hardware cost by re-utilizing the idle BCH encoder at that time. Moreover, Peterson's algorithm has been employed in the second stage of BCH decoding in [11], that its complexity increases significantly for larger $t$. In contrast, our decoder utilizes the FIBM algorithm, which not only saves a considerable amount of silicon area but also reduces power consumption. What's more, our proposed bit-parallel polynomial basis (PB) multipliers which requires no XOR gates and provide an area-efficient architecture compared to other existing multipliers in the literature resulting in the regularity and modularity of VLSI implementation of finite field multipliers.

## VII. CONCLUSION

In this study, we present a novel fault detection architecture that boasts a remarkable efficiency in terms of area and delay. This architecture is based on the BCH error correction code, utilized in a new parallel polynomial basis multipliers that are endowed with the unique ability to accurately rectify multiple errors. We also introduce the optimized Berlekamp-Rumsey-Solomon (BRS) algorithm [19], [22], in conjunction with the Chien-search algorithm, which expediently detecting the location of errors in the output of the multiplier. Our proposed scheme is subjected to a rigorous complexity analysis, which unequivocally demonstrates that it has a lower complexity in terms of area, delay, and power compared to other well-known techniques. This feature is highly desirable in applications that are constrained, such as smart cards, IoT devices, and implantable medical devices.